\documentclass{ase}

\begin{document}

\begin{center}
{\huge A New Metric for Quality of Network Community Structure}


\begin{multicols}{3} 
{Mingming Chen}{\\}{Department of Computer Science\\Rensselaer Polytechnic Institute\\110 8th Street\\Troy, New York 12180\\Email: chenm8@rpi.edu\\ \columnbreak}
{Tommy Nguyen}{\\}{Department of Computer Science\\Rensselaer Polytechnic Institute\\110 8th Street\\Troy, New York 12180\\Email: nguyet11@rpi.edu \\ \columnbreak}
{Boleslaw K. Szymanski}{\\}{Department of Computer Science\\Rensselaer Polytechnic Institute\\110 8th Street\\Troy, New York 12180\\Email: szymab@rpi.edu}\\
\end{multicols}
\end{center}

\begin{multicols*}{2}

\begin{abstract}
Modularity is widely used to effectively measure the strength of the community structure found by community detection algorithms. However, modularity maximization suffers from two opposite yet coexisting problems: in some cases, it tends to favor small communities over large ones while in others, large communities over small ones. The latter tendency is known in the literature as the resolution limit problem. To address them, we propose to modify modularity by subtracting from it the fraction of edges connecting nodes of different communities and by including community density into modularity. We refer to the modified metric as \textit{Modularity Density} and we demonstrate that it indeed resolves both problems mentioned above. We describe the motivation for introducing this metric by using intuitively clear and simple examples. We also prove that this new metric solves the resolution limit problem. Finally, we discuss the results of applying this metric, modularity, and several other popular community quality metrics to two real dynamic networks. The results imply that \textit{Modularity Density} is consistent with all the community quality measurements but not modularity, which suggests that \textit{Modularity Density} is an improved measurement of the community quality compared to modularity.
\end{abstract}

\section{Introduction}
Communities are the basic structures in sociology in general and in social networks in particular. They have been intensively researched for more than a half of the century \cite{HumanCommunities}. Community in sociology usually refers to a social unit whose members share common values and the identity of the members as well as their degree of cohesiveness depend on individuals' social and cognitive factors such as beliefs, preferences, or needs. The ubiquity of the Internet and social media eliminated spatial limitations on community geographical range, enabling on-line communities to link people regardless of their physical location. The newly arising computational sociology relies on computationally intensive methods to analyze and model social phenomena \cite{ComSociology}, including communities and their detection.

Analysis of social networks became one of the basic tools of sociology \cite{SocNetMeth_App} and has been used for linking micro and macro levels of sociological theory. The classical example of the approach is presented in \cite{WeakTies} that elaborated the macro implications of one aspect of
small-scale interaction, the strength of dyadic ties. Moreover, a lot of commercial applications, such as digital marketing, behavioral targeting, and user preference mining, rely heavily on community analysis. With the rapid growth of large-scale on-line social networks, e.g., Facebook connected a billion users in 2012, there is a high demand for efficient community detection algorithms that will be able to handle their evolution growth. Communities in on-line social networks are discovered by analyzing the observed and often recorded on-line interactions between people.

In computational sociology, communities are defined as groups of nodes in a social network within which connections are denser than between them \cite{UWDModularity}. This definition has been found useful also in other type of networks, and community detection became one of the fundamental issues in network science. Community detection has been shown to reveal latent yet meaningful structure not only for groups in online and contact-based social networks, but also in groups of customers with similar interests in online retailer user networks, groups of scientists in interdisciplinary collaboration networks, and in biology in functional modules in protein-protein interaction networks etc. \cite{CommunityReport}. Since in most applications the real communities are not known (often due to the cost of establishing ground truth in large on-line social  networks), there is a need for developing reliable metrics to evaluate detected communities, so these metrics can be used to rank the quality of community structures discovered by different community detection algorithms. Such metrics can also be used to develop novel community algorithms that iteratively attempt to improve the metrics by merging or splitting the given network community structure.

In the last decade, the most popular community detection method, proposed by Newman \cite{NewmanGreedy}, has been to maximize the quality metric known as modularity \cite{UWDModularity,PNASModularity} over all the possible partitions of a network. This metric measures the difference (relative to the total number of edges) between the actual and expected (in a randomized graph with the same number of nodes and the same degree distribution) number of edges within a given community. It is widely used to measure the strength of the community structures detected by the community detection algorithms. However, modularity maximization has two opposite yet concurrent problems. In some cases, it tends to split large communities into smaller communities. In other cases, it tends to form large communities by merging communities that are smaller than a certain threshold which depends on the total number of edges in the network and on the degree of inter-connectivity between the communities. The latter problem is known as the resolution limit problem \cite{ResolutionLimit}.

To solve these two problems simultaneously, we propose a new community quality metric, that we termed \textit{Modularity Density}, as an alternative to modularity. First, we show modularity decreased by \textit{Split Penalty}, defined as the fraction of edges that connect nodes of different communities, solves the problem of favoring small communities. Next, we demonstrate that including community density into modularity addresses the problem of favoring large communities. We refer to the resulting metric as \textit{Modularity Density}.

We formally prove that \textit{Modularity Density} could resolve the resolution limit problem. We also discuss our experiments with this metric, modularity, and other popular community quality metrics, including the number of \textit{Intra-edges}, \textit{Contraction}, the number of \textit{Inter-edges}, \textit{Expansion}, and \textit{Conductance} \cite{CommunityMetrics}, on two real dynamic networks. The results show that \textit{Modularity Density} is different from original modularity, but consistent with all those community quality measurements, which implies that \textit{Modularity Density} is effective in measuring the community quality of networks.

The rest of the paper is organized as follows. First, in Section~\ref{sec:related_work} we discuss some related works. Then, we briefly introduce modularity and illustrate our motivation to propose the new metric with examples in Section~\ref{sec:modularity_density}. Section~\ref{sec:evaluation} presents the formal proofs and the experiments that demonstrate \textit{Modularity Density} solves the two problems of modularity simultaneously. Finally, we conclude and discuss the future work in Section~\ref{sec:conclusion}.

\section{Related Work}
\label{sec:related_work}
Community detection in complex networks has received a considerable amount of attention in the last years. Numerous techniques have been developed for both efficient and effective community detection, including Modularity Optimization \cite{NewmanGreedy,PNASModularity,ModularityLargeNet,Agglomeration,Spectral,Extremal,SimulatedAnnealing}, Clique Percolation \cite{CPM,CPMw}, Local Expansion \cite{LocalExpansionRPI,LFM,EAGLE}, Fuzzy Clustering \cite{ZhangFuzzy,NMFFuzzy}, Link Partitioning \cite{LinkPartition}, and Label Propagation \cite{LPA,SLPA2011,SLPA2012}. The above algorithms are designed to detect communities on static networks. However, networks, such as Internet and online social networks, are usually dynamic, with changes arriving as a stream. Thus, a large number of algorithms were proposed to cope with community detection on dynamically evolving networks, such as LabelRankT \cite{LabelRankT} and Estrangement \cite{Estrangement}. LabelRankT \cite{LabelRankT} detects communities in large-scale dynamic networks through stabilized label propagation. Estrangement \cite{Estrangement} detects temporal communities by maximizing modularity in a snapshot subject to a constraint on the estrangement from the partition in the previous snapshot.

In addition to the development of algorithms for community detection, several metrics for evaluating the quality of community structure have been introduced. The most popular and widely used is modularity \cite{UWDModularity,PNASModularity}. It is defined as the difference (relative to the total number of edges) between the actual and expected (in a randomized graph with the same number of nodes and the same degree sequence) number of edges inside a given community. Although initially defined for unweighted and undirected networks, the definition of modularity has been subsequently extended to capture community structure in weighted networks \cite{WeightedModularity} and then in directed networks \cite{DirectedModularity}.

\begin{figure*}[!t]
\centering
\setlength{\belowcaptionskip}{-1em}
\subfigure[Two very well separated communities.]{
\label{fig_sp_examples:subfig:a}
\includegraphics[scale=0.53]{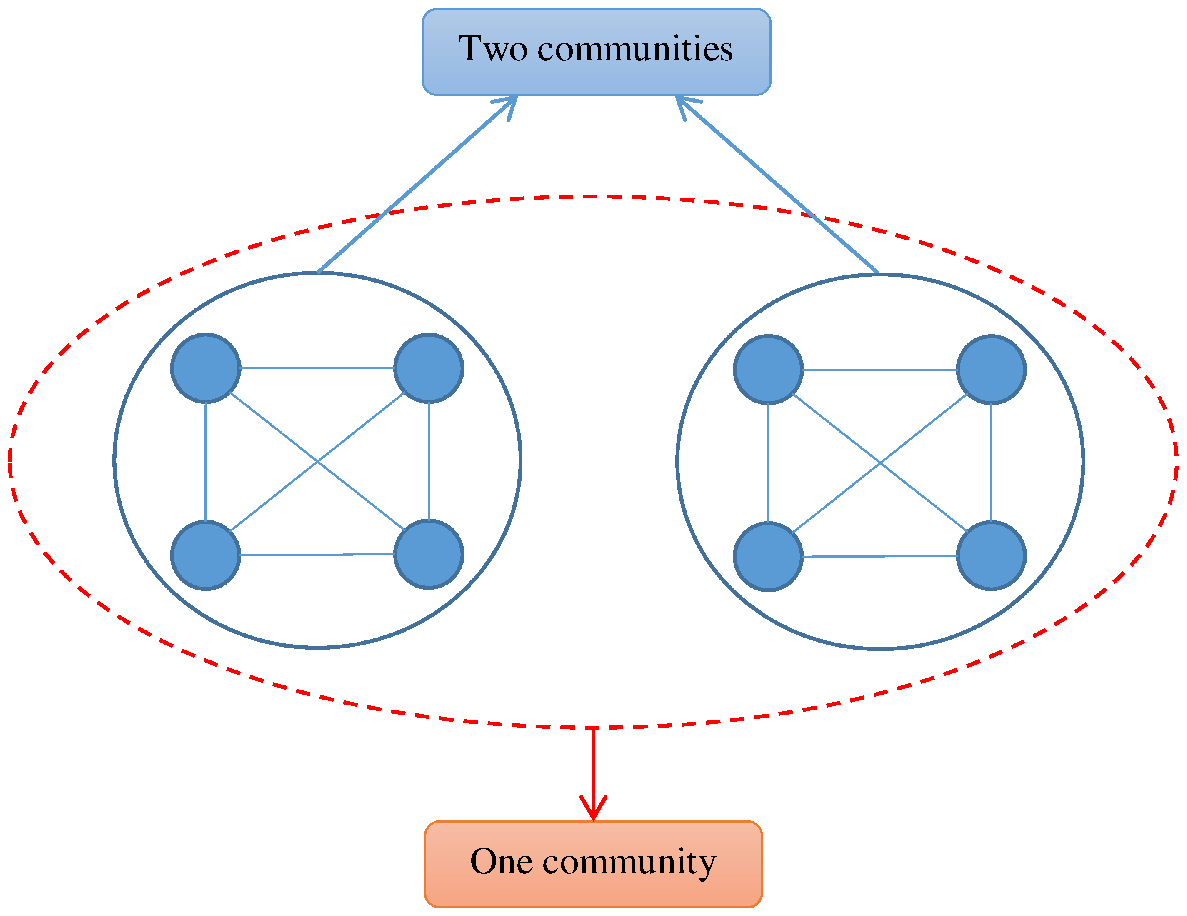}
}
\hspace{5em}
\subfigure[Two well separated communities.]{
\label{fig_sp_examples:subfig:b}
\includegraphics[scale=0.52]{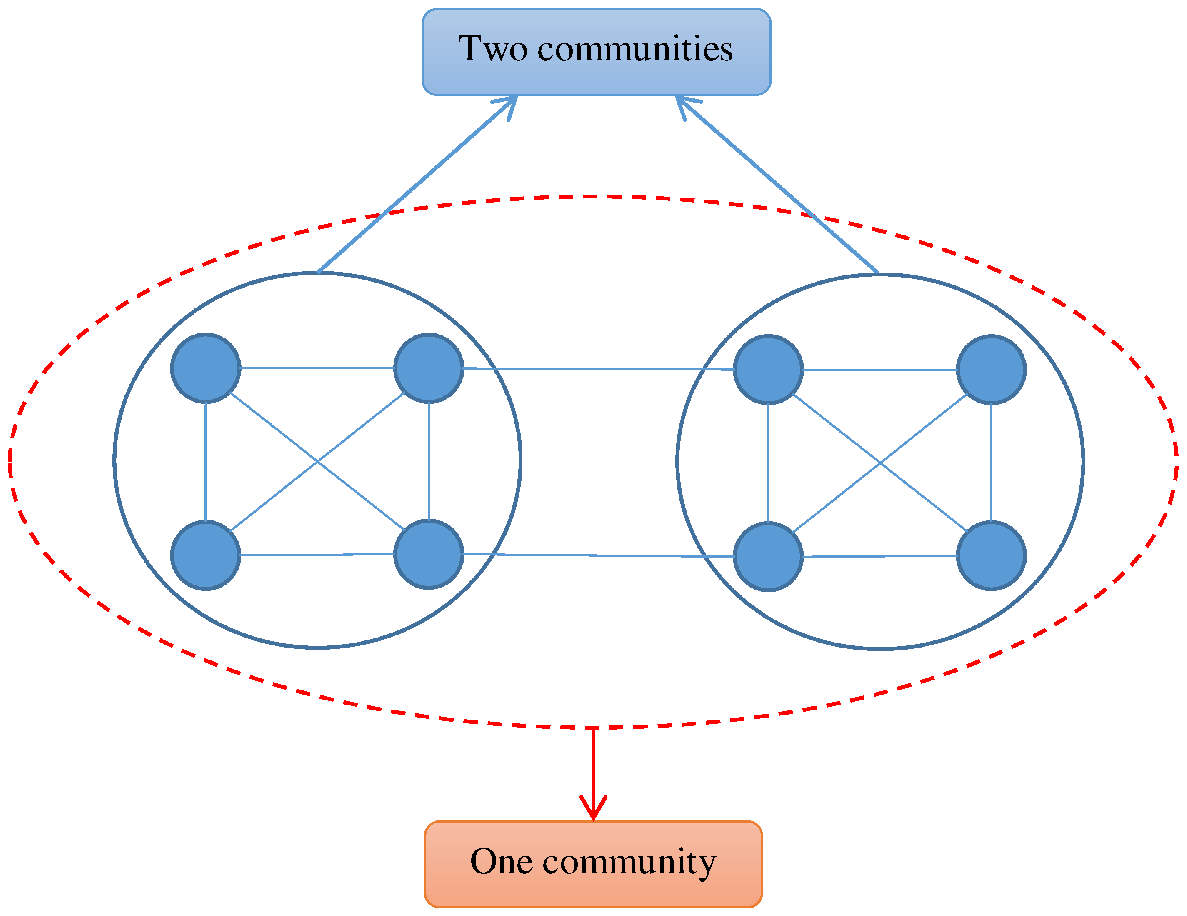}
}
\vspace{0.1em}
\subfigure[Two weakly connected communities.]{
\label{fig_sp_examples:subfig:c}
\includegraphics[scale=0.52]{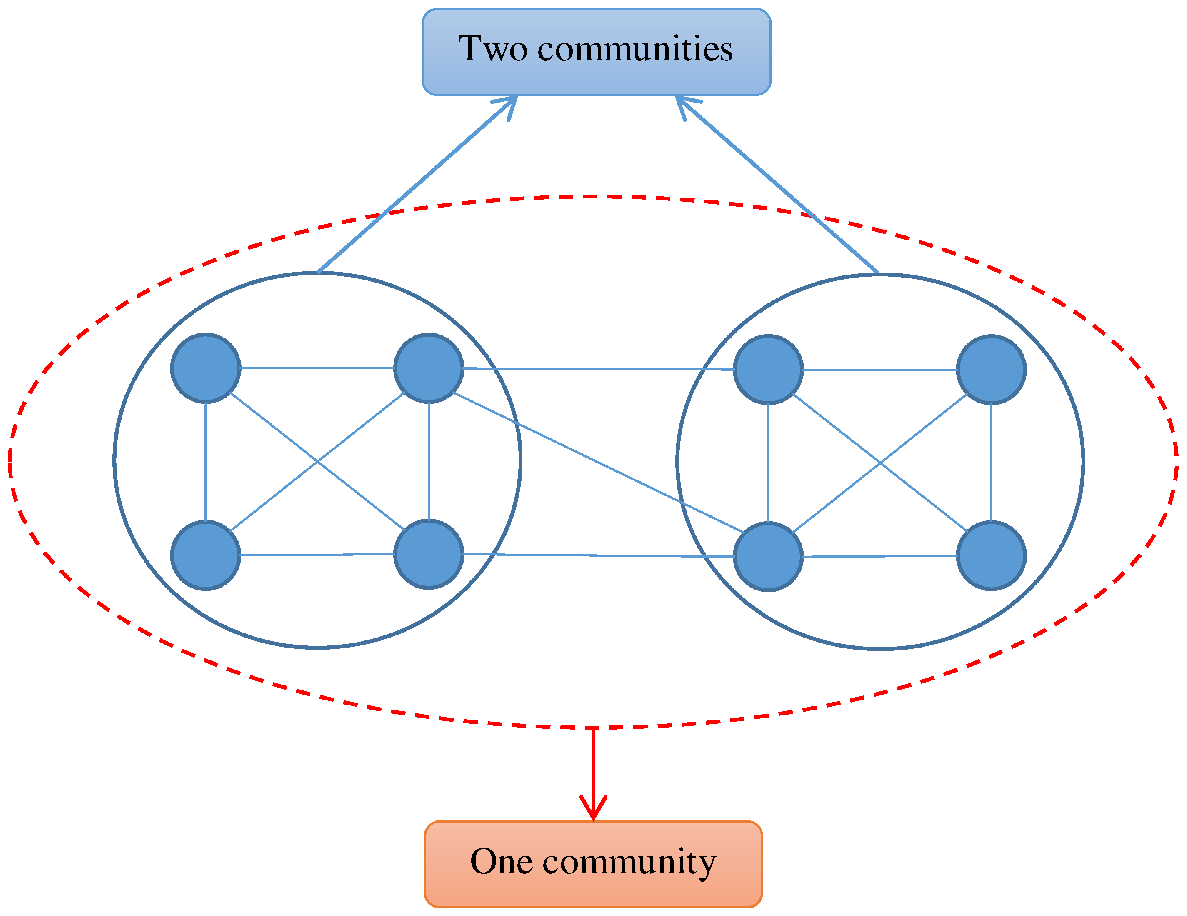}
}
\hspace{5em}
\subfigure[Ambiguity between one and two communities.]{
\label{fig_sp_examples:subfig:d}
\includegraphics[scale=0.52]{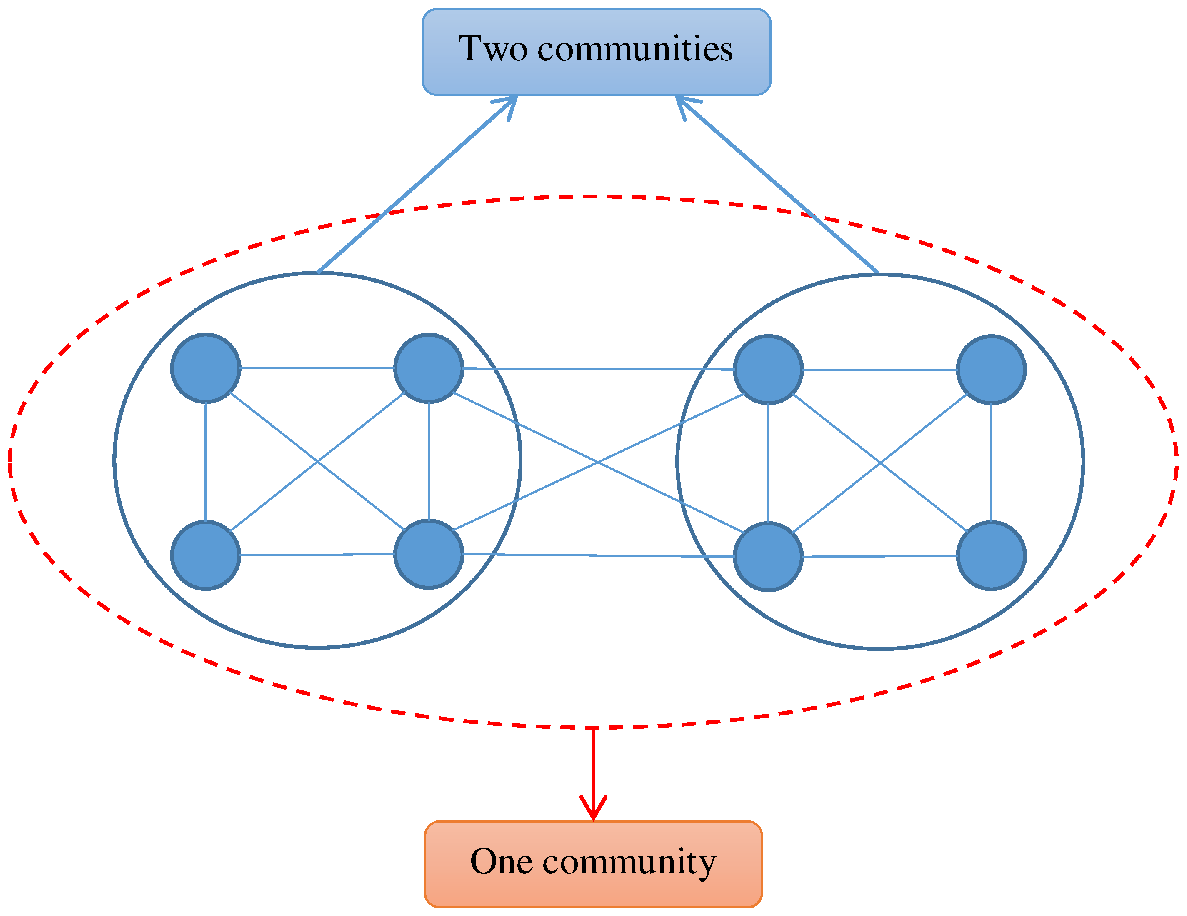}
}
\vspace{0.1em}
\subfigure[One well connected community.]{
\label{fig_sp_examples:subfig:e}
\includegraphics[scale=0.52]{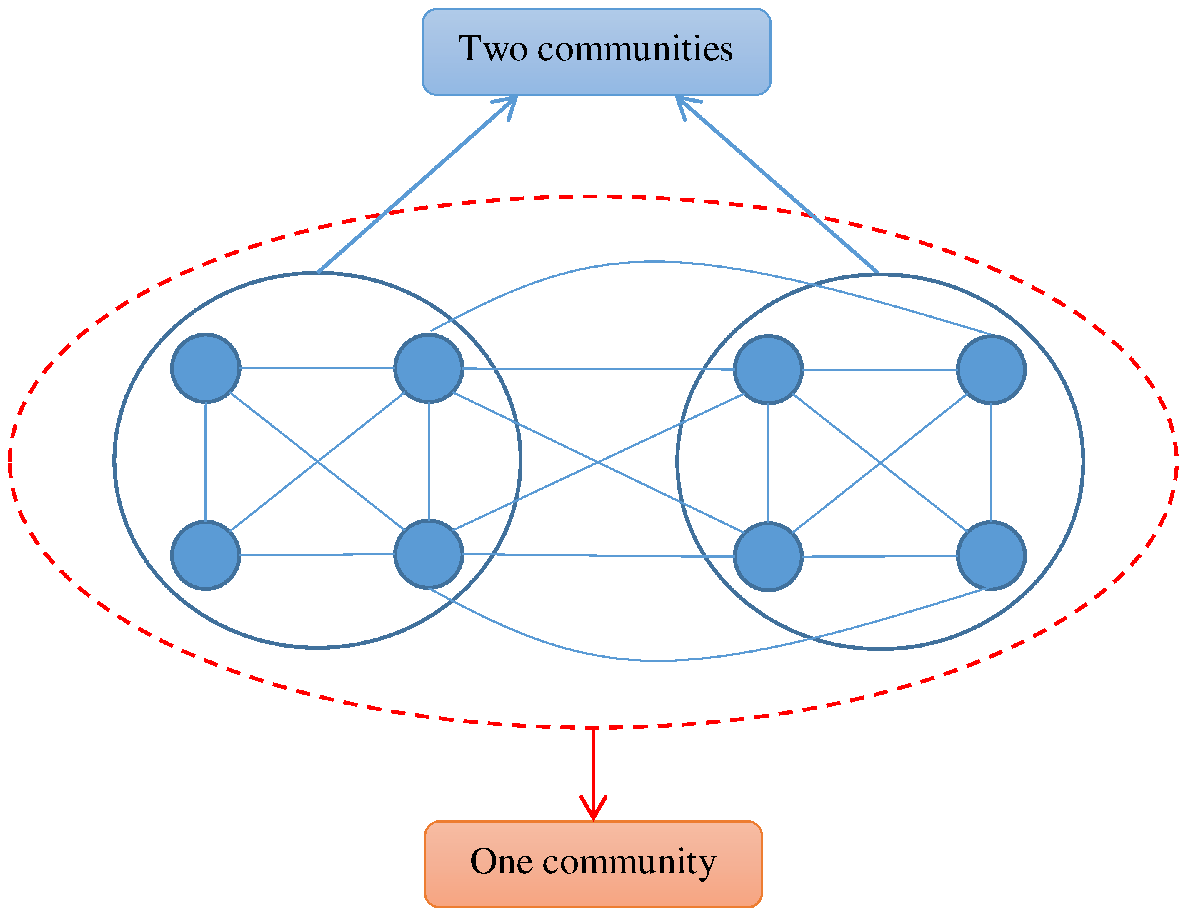}
}
\hspace{5em}
\subfigure[One very well connected community.]{
\label{fig_sp_examples:subfig:f}
\includegraphics[scale=0.52]{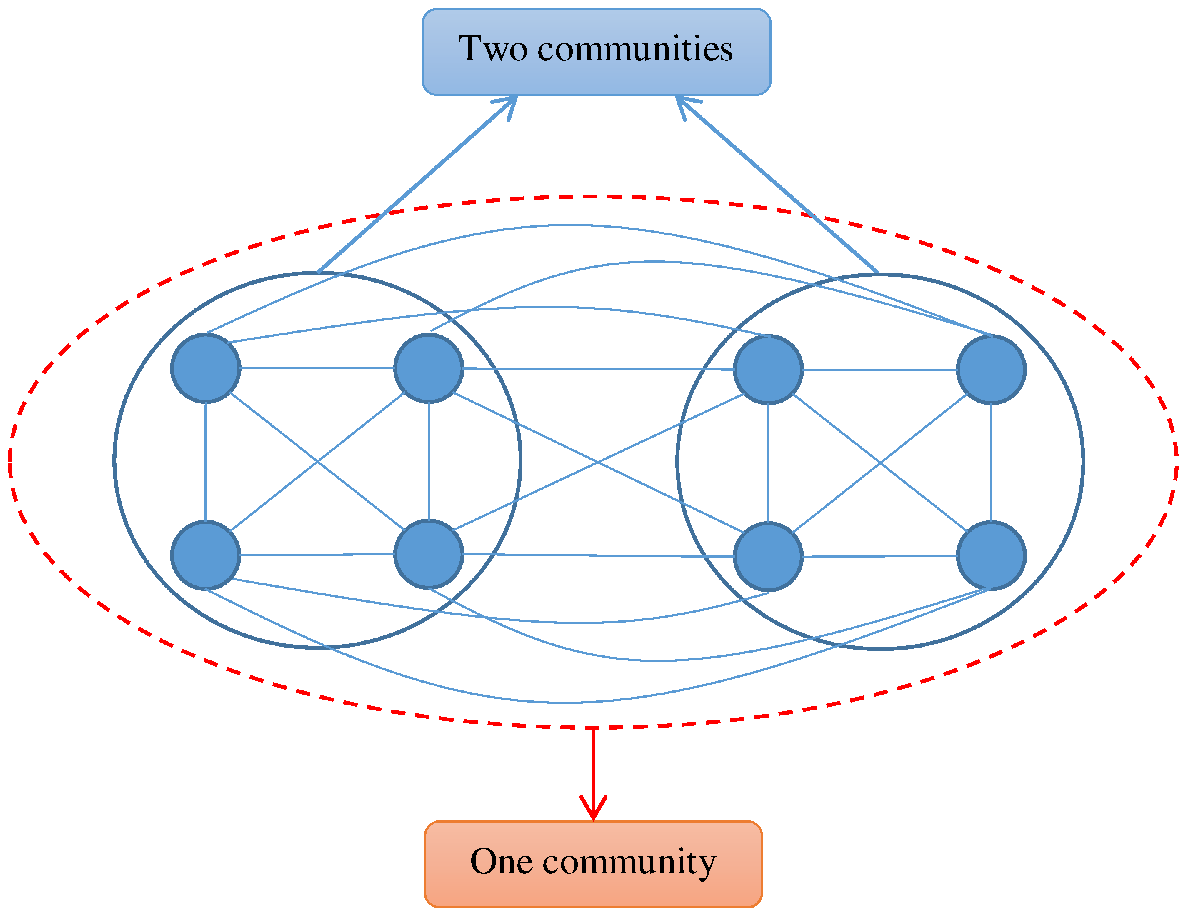}
}
\vspace{-1.5em}
\centering
\caption{Six simple network examples that have two different community structures, one with a single big community containing all eight nodes and the other with the two small communities each containing four different nodes.}
\label{fig_sp_examples}
\vspace{0.6em}
\end{figure*}

However, recently, Fortunato and Barth\'{e}lemy \cite{ResolutionLimit} presented a resolution limit problem of modularity, essence of which is that optimizing modularity will not find communities smaller than a threshold size, or weight \cite{ResolutionLimitWeight}. This threshold depends on the total number, or total weight, of edges in the network and on the degree of interconnectedness between the communities. Moreover, Good et al. \cite{ExtremeDegeneracy} shown that the range of modularity values computed over all possible partitions of a graph has a structure in which the maximum modularity partition is typically concealed among an exponentially large (in terms of the graph size) number of structurally dissimilar, high-modularity partitions. To address this resolution limit problem, multi-resolution versions of modularity \cite{ResolutionLimitParam1,ResolutionLimitParam2} were proposed to allow researchers to specify a tunable target resolution limit parameter and identify communities on that scale. Typically, it is not clear how to choose the correct value for this parameter. Furthermore, Lancichinetti and Fortunato \cite{Modularity2Problems} stated that even those multi-resolution versions of modularity as well as its original version are not only inclined to merge the smallest well-formed communities but also to split the largest well-formed communities. In contrast, the \textit{Modularity Density} metric we propose here solves those two problems of modularity without the trouble of specifying any particular parameter.

\begin{table*}[!t]
\caption{Metric values of the example: Two very well separated communities.}
\label{tab:simple_example1}
\centering
\begin{tabular}{l||c|c|c|c}
\hline \hline
     & \textit{~Modularity} ($Q$)~ & \textit{Split Penalty} ($SP$) & ~~~~~~$Q_s$~~~~~~ & ~~~~~~$Q_{ds}$~~~~~~ \\
\hline
     Two communities & 0.5 & 0 & 0.5 & 0.5 \\
\hline
     One community & 0 & 0 & 0 & 0.245 \\
\hline \hline
\end{tabular}
\vspace{-1.3em}
\end{table*}

\begin{table*}[!t]
\caption{Metric values of the example: Two well separated communities.}
\label{tab:simple_example2}
\centering
\begin{tabular}{l||c|c|c|c}
\hline \hline
     & \textit{~Modularity} ($Q$)~ & \textit{Split Penalty} ($SP$) & ~~~~~~$Q_s$~~~~~~ & ~~~~~~$Q_{ds}$~~~~~~ \\
\hline
     Two communities & 0.357 & 0.143 & 0.214 & 0.339 \\
\hline
     One community & 0 & 0 & 0 & 0.25 \\
\hline \hline
\end{tabular}
\vspace{-1.3em}
\end{table*}

\begin{table*}[!t]
\caption{Metric values of the example: Two weakly connected communities.}
\label{tab:simple_example3}
\centering
\begin{tabular}{l||c|c|c|c}
\hline \hline
     & \textit{~Modularity} ($Q$)~ & \textit{Split Penalty} ($SP$) & ~~~~~~$Q_s$~~~~~~ & ~~~~~~$Q_{ds}$~~~~~~ \\
\hline
     Two communities & 0.3 & 0.2 & 0.1 & 0.263 \\
\hline
     One community & 0 & 0 & 0 & 0.249 \\
\hline \hline
\end{tabular}
\vspace{-0.4em}
\end{table*}

\section{Modularity Density}
\label{sec:modularity_density}
In this section, we first formally introduce Newman's definition of modularity and then illustrate the motivation for modifying modularity with several simple network examples. Next, we propose a new community quality metric, called \textit{Modularity Density}, as an alternative to modularity by combining modularity with \textit{Split Penalty} and community density to avoid the two coexisting problems of modularity. Finally, we define \textit{Modularity Density} for different kinds of networks, including unweighted and undirected networks, weighted networks, and directed networks, based on the corresponding formulas of modularity.

\subsection{Newman's Modularity}
\label{subsec:modularity}
Modularity \cite{UWDModularity,PNASModularity} for unweighted and undirected networks is defined as the ratio of difference between the actual and expected (in a randomized graph with the same number of nodes and the same degree sequence) number of edges within the community. For the given community partition of a network $G=(V,E)$ with $|E|$ edges, modularity ($Q$) \cite{UWDModularity} is given by
\begin{equation}
\label{eq:uwdmodularity}
Q=\sum_{c_i \in C} \left[\frac{|E_{c_i}^{in}|}{|E|}-\left(\frac{2|E_{c_i}^{in}|+|E_{c_i}^{out}|}{2|E|}\right)^2\right],
\end{equation}
where $C$ is the set of all the communities, $c_i$ is a specific community in $C$, $|E_{c_i}^{in}|$ is the number of edges between nodes within community $c_i$, and $|E_{c_i}^{out}|$ is the number of edges from the nodes in community $c_i$ to the nodes outside $c_i$.


The definition of modularity \cite{WeightedModularity} for the weighted networks has precisely the same formula, Equation~(\ref{eq:uwdmodularity}), as for the unweighted and undirected networks. However, for weighted networks, $|E|$ is the sum of the weights of all the edges in the network, $|E_{c_i}^{in}|$ is the sum of the weights of the edges between nodes within community $c_i$, and $|E_{c_i}^{out}|$ is the sum of the weights of the edges from the nodes in community $c_i$ to the nodes outside $c_i$.


The formula of modularity for directed networks \cite{DirectedModularity} is as follows
\begin{equation}
\label{eq:directedmodularity}
Q=\sum_{c_i \in C}\left[\frac{|E_{c_i}^{in}|}{|E|}-\frac{(|E_{c_i}^{in}|+|E_{out,c_i}|)(|E_{c_i}^{in}|+|E_{c_i,out}|)}{|E|^2}\right],
\end{equation}
where $|E_{out,c_i}|$ is the number of edges from the nodes outside $c_i$ to the nodes in $c_i$ and $|E_{c_i,out}|$ is the number of edges from the nodes in $c_i$ to the nodes outside $c_i$. For undirected networks, it is clear that $|E_{out,c_i}|=|E_{c_i,out}|=|E_{c_i}^{out}|$ and thus the directed modularity is reduced to undirected modularity.

\subsection{Motivation for Introducing Split Penalty}
\label{subsec:motivation_sp}
In this subsection, we demonstrate the motivation for introducing \textit{Split Penalty} into modularity by using seven intuitively clear and simple network examples, six of which are presented in Figure~\ref{fig_sp_examples}. The seventh example is a complete graph with eight nodes and one big community containing all eight nodes while the alternative partition consists of the two small communities each containing four different nodes. We could easily judge that for the first, second, and the third examples, the community structure with two small communities is better than the community structure in which they are merged together. For the fourth example, the two different community structures are nearly of the same quality. However, for the fifth, sixth, and the seventh examples, the community structure with one big community is of better quality than the alternative.

\begin{table*}[!t]
\caption{Metric values of the example: Ambiguity between one and two communities.}
\label{tab:simple_example4}
\centering
\begin{tabular}{l||c|c|c|c}
\hline \hline
     & \textit{~Modularity} ($Q$)~ & \textit{Split Penalty} ($SP$) & ~~~~~~$Q_s$~~~~~~ & ~~~~~~$Q_{ds}$~~~~~~ \\
\hline
     Two communities & 0.25 & 0.25 & 0 & 0.188 \\
\hline
     One community & 0 & 0 & 0 & 0.245 \\
\hline \hline
\end{tabular}
\vspace{-1.3em}
\end{table*}

\begin{table*}[!t]
\caption{Metric values of the example: One well connected community.}
\label{tab:simple_example5}
\centering
\begin{tabular}{l||c|c|c|c}
\hline \hline
     & \textit{~Modularity} ($Q$)~ & \textit{Split Penalty} ($SP$) & ~~~~~~$Q_s$~~~~~~ & ~~~~~~$Q_{ds}$~~~~~~ \\
\hline
     Two communities & 0.167 & 0.333 & -0.167 & 0.0417 \\
\hline
     One community & 0 & 0 & 0 & 0.23 \\
\hline \hline
\end{tabular}
\vspace{-1.3em}
\end{table*}

\begin{table*}[!t]
\caption{Metric values of the example: One very well connected community.}
\label{tab:simple_example6}
\centering
\begin{tabular}{l||c|c|c|c}
\hline \hline
     & \textit{~Modularity} ($Q$)~ & \textit{Split Penalty} ($SP$) & ~~~~~~$Q_s$~~~~~~ & ~~~~~~$Q_{ds}$~~~~~~ \\
\hline
     Two communities & 0.0455 & 0.455 & -0.409 & -0.239 \\
\hline
     One community & 0 & 0 & 0 & 0.168 \\
\hline \hline
\end{tabular}
\vspace{-1.3em}
\end{table*}

\begin{table*}[!t]
\caption{Metric values of the example: One Complete Graph.}
\label{tab:simple_example7}
\centering
\begin{tabular}{l||c|c|c|c}
\hline \hline
     & \textit{~Modularity} ($Q$)~ & \textit{Split Penalty} ($SP$) & ~~~~~~$Q_s$~~~~~~ & ~~~~~~$Q_{ds}$~~~~~~ \\
\hline
     Two communities & -0.0714 & 0.571 & -0.643 & -0.643 \\
\hline
     One community & 0 & 0 & 0 & 0 \\
\hline \hline
\end{tabular}
\vspace{-0.4em}
\end{table*}

Tables~\ref{tab:simple_example1}-\ref{tab:simple_example7} show the metric values of the seven network examples described above. Tables~\ref{tab:simple_example1}-\ref{tab:simple_example3}, and Table~\ref{tab:simple_example7} demonstrate that modularity succeeds in measuring the quality of the two different community structures in those four examples. However, from Tables~\ref{tab:simple_example4}-\ref{tab:simple_example6}, we could observe that modularity actually fails to measure the community quality of those three examples because it implies that the community structure with two small communities is better. In contrast, for the fifth and the sixth examples, the community structure with one big community is of better quality. Yet, in this case modularity gives preference to the community structure with two separated small communities, demonstrating that modularity has the problem of favoring small communities.

To address the drawback of favoring small communities,  we propose that the quality of the community structure should take into account the edges between different communities. We introduce \textit{Modularity with Split Penalty} ($Q_s$) by subtracting from modularity the \textit{Split Penalty} ($SP$) which is the fraction of edges that connect nodes of different communities. More formally,
\begin{equation}
\label{eq:modularity_sp}
Q_s=Q-SP.
\end{equation}
The intuition here is clear. Modularity measures the positive effect of grouping nodes together in terms of taking into account existing edges between nodes while \textit{Split Penalty} measures the negative effect of ignoring edges joining members of different communities. Enlarging community eliminates some \textit{Split Penalty} but if there are only a few edges across current partition, modularity of the merged community could be lower, negating the benefit of merging. Splitting a community into two or more communities introduces some \textit{Split Penalty} but if there are only a few edges between those separated communities, an increase of modularity can make such splitting beneficial. Tables \ref{tab:simple_example1}-\ref{tab:simple_example7} demonstrate that $Q_s$ can correctly measure the quality of the community structures of all seven network examples.

\begin{figure*}[!t]
\centering
\setlength{\belowcaptionskip}{-1em}
\subfigure[Two clique communities.]{
\label{clique_tree:subfig:a}
\includegraphics[scale=0.62]{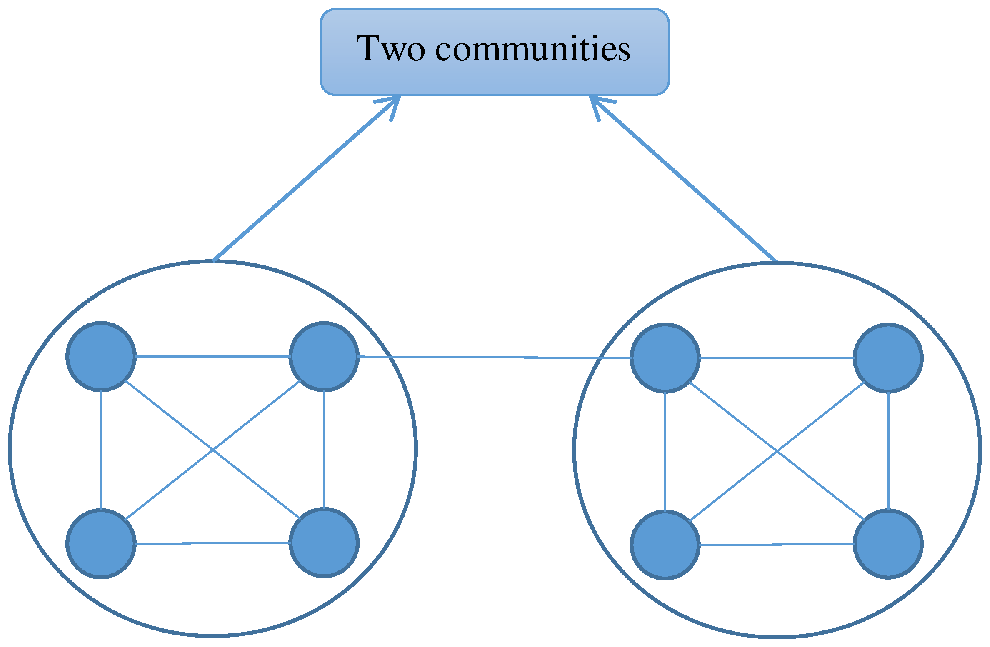}
}
\hspace{4em}
\subfigure[Two tree communities.]{
\label{clique_tree:subfig:b}
\includegraphics[scale=0.62]{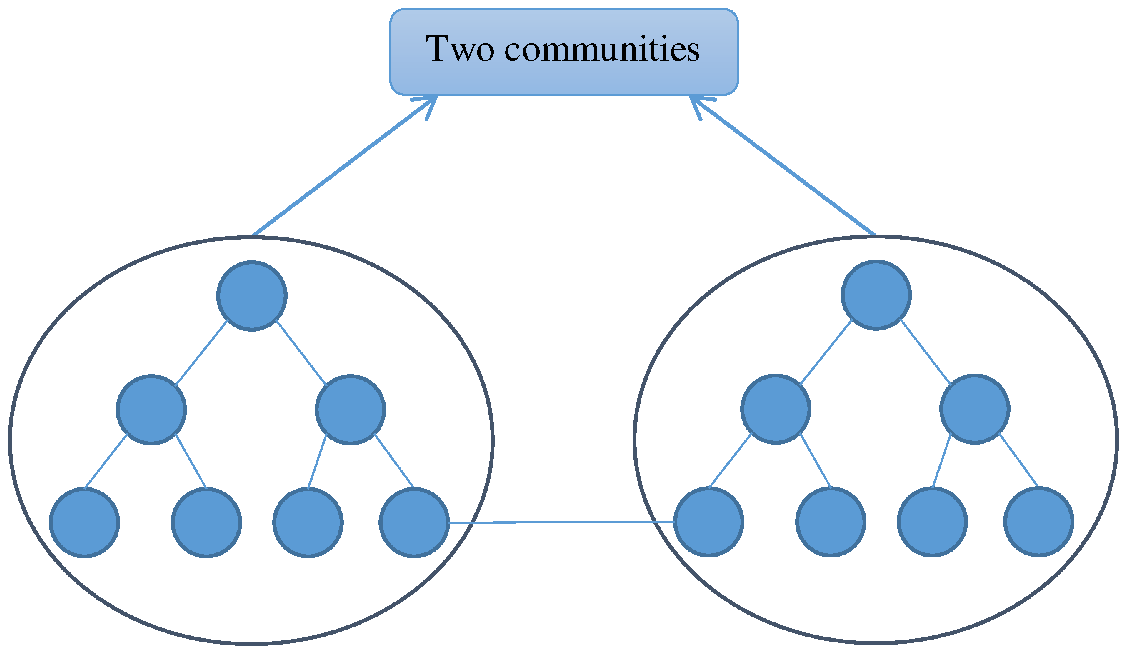}
}
\vspace{-1.4em}
\centering
\caption{Two simple network examples with the left one containing two clique communities and the right one containing two tree communities. Also, there are six edges within all four communities, but the number of nodes is different in clique and tree communities.}
\label{clique_tree}
\end{figure*}

\begin{table*}[!t]
\caption{Metric values of the example: two clique communities vs two tree communities.}
\label{tab:simple_example8}
\centering
\begin{tabular}{l||c|c|c|c}
\hline \hline
     & \textit{~Modularity} ($Q$)~ & \textit{Split Penalty} ($SP$) & ~~~~~~$Q_s$~~~~~~ & ~~~~~~$Q_{ds}$~~~~~~ \\
\hline
     Two clique communities & 0.4231 & 0.07692 & 0.3462 & 0.4183 \\
\hline
     Two tree communities & 0.4231 & 0.07692 & 0.3462 & 0.2214 \\
\hline \hline
\end{tabular}
\end{table*}

\subsection{Modularity with Split Penalty}
\label{subsec:modularity_sp}
In this subsection, we extend the formula of $Q_s$ to different kinds of networks, such as unweighted and undirected networks, weighted networks, and directed networks, based on the corresponding formulas of modularity presented in Subsection~\ref{sec:modularity_density}-\ref{subsec:modularity}.

From Subsection~\ref{sec:modularity_density}-\ref{subsec:motivation_sp}, we know that \textit{Split Penalty} ($SP$) is the fraction of edges that connect nodes of different communities. Thus, for undirected networks, no matter unweighted or weighted, \textit{Split Penalty} is defined as
\begin{equation}
\label{eq:uwdsp}
SP=\sum_{c_i \in C}\biggl[\sum_{\substack{c_j \in C \\ c_j \ne c_i}}\frac{|E_{c_i,c_j}|}{2|E|}\biggr].
\end{equation}
where $|E_{c_i,c_j}|$ is the number of edges from community $c_i$ to community $c_j$ for unweighted networks or the sum of the weights of the edges from community $c_i$ to community $c_j$ for weighted networks.
For directed networks, \textit{Split Penalty} is given by
\begin{equation}
\label{eq:directedsp}
SP=\sum_{c_i \in C}\biggl[\sum_{\substack{c_j \in C \\ c_j \ne c_i}}\frac{|E_{c_i,c_j}|}{|E|}\biggr].
\end{equation}
It can be seen that for each community, the \textit{Split Penalty} only takes into account the outgoing edges from this community to the rest of the network but not the incoming edges from the rest of the network to this community. It is reasonable to use only outgoing edges, because in a sense those are friendships of community members. Incoming edges may not be apparent. Moreover, considering both outgoing and incoming edges would only double the value of \textit{Split Penalty} because the incoming edges of a community are the outgoing edges of other communities.

Therefore, for undirected networks, both unweighted and weighted, from Equations~(\ref{eq:uwdmodularity}),~(\ref{eq:modularity_sp}), and~(\ref{eq:uwdsp}), $Q_s$ is defined as
%
\begin{equation}
\label{eq:uwdqs}
\begin{split}
&Q_s=Q-SP \\
&=\sum_{c_i \in C} \left[\frac{|E_{c_i}^{in}|}{|E|}-\left(\frac{2|E_{c_i}^{in}|+|E_{c_i}^{out}|}{2|E|}\right)^2-\sum_{\substack{c_j \in C \\ c_j \ne c_i}}\frac{|E_{c_i,c_j}|}{2|E|}\right].
\end{split}
\end{equation}
For directed networks, using Equations~(\ref{eq:directedmodularity}),~(\ref{eq:modularity_sp}), and~(\ref{eq:directedsp}), $Q_s$ can be expressed as
%
\begin{equation}
\label{eq:directedqs}
\begin{split}
&Q_s=Q-SP \\
&=\sum_{c_i \in C} \biggl[\frac{|E_{c_i}^{in}|}{|E|}-\frac{(|E_{c_i}^{in}|+|E_{out,c_i}|)(|E_{c_i}^{in}|+|E_{c_i,out}|)}{|E|^2} \\
&~~~~~~~~~~~~~~~-\sum_{\substack{c_j \in C \\ c_j \ne c_i}}\frac{|E_{c_i,c_j}|}{|E|}\biggr].
\end{split}
\end{equation}

\begin{figure*}[!t]
\centering
\setlength{\belowcaptionskip}{-1em}
\includegraphics[scale=0.52]{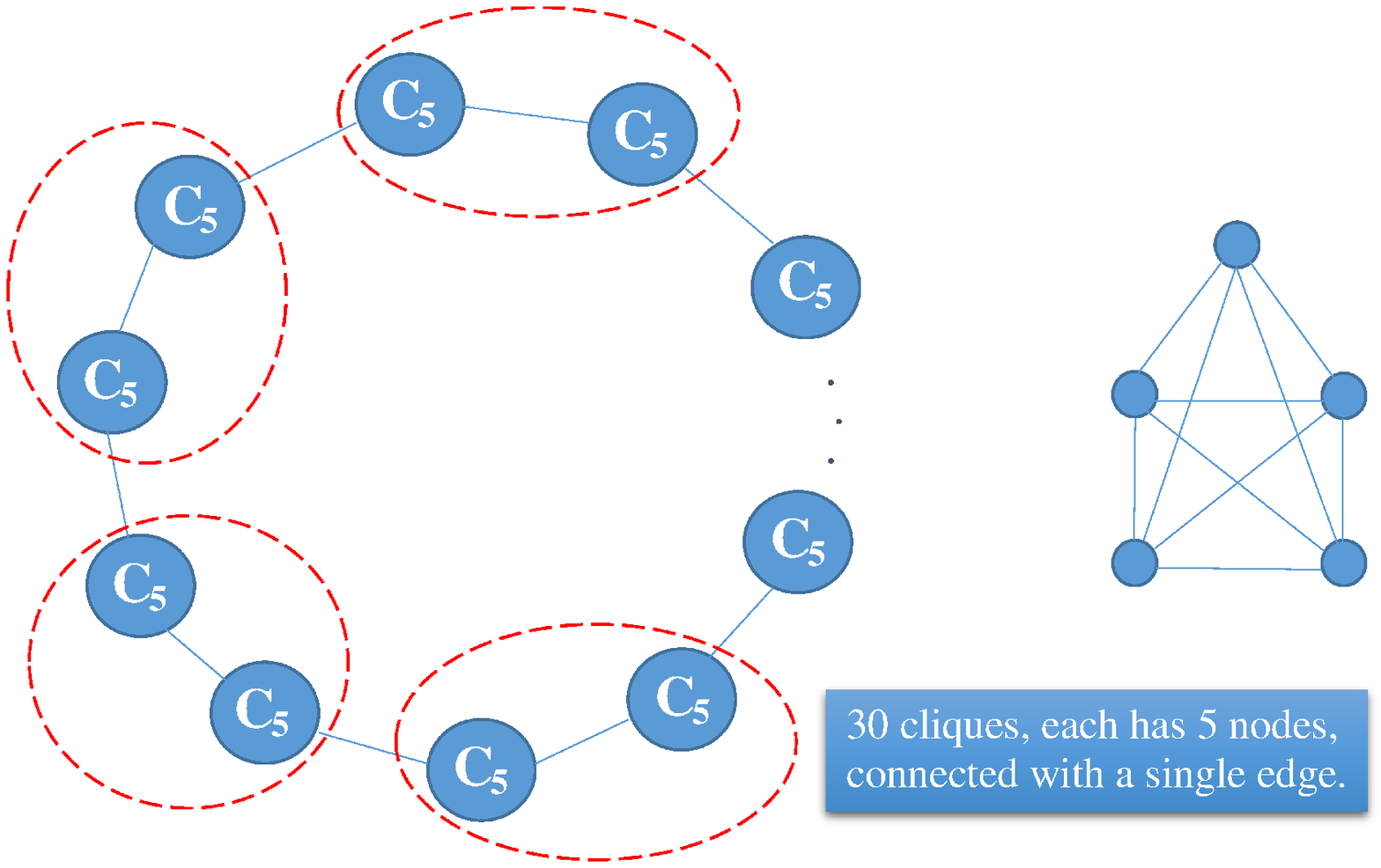}
\vspace{-1em}
\centering
\caption{A ring network example made out of thirty identical cliques, each having five nodes and connected by single edges.}
\label{30_cliques}
\vspace{0.2em}
\end{figure*}

\begin{table*}[!t]
\caption{Metric values of the example: a ring of thirty cliques, each having five nodes and connected by single edges.}
\label{tab:simple_example9}
\centering
\begin{tabular}{l||c|c|c|c}
\hline \hline
     & \textit{~Modularity} ($Q$)~ & \textit{Split Penalty} ($SP$) & ~~~~~~$Q_s$~~~~~~ & ~~~~~~$Q_{ds}$~~~~~~ \\
\hline
     Thirty communities & 0.8758 & 0.09091 & 0.7848 & 0.8721 \\
\hline
     Fifteen communities & 0.8879 & 0.04545 & 0.8424 & 0.4305 \\
\hline \hline
\end{tabular}
\vspace{-0.4em}
\end{table*}

\subsection{Motivation for Introducing Community Density}
\label{subsec:motivation_cd}
Modularity and also $Q_s$ have two shortcomings. First, they are independent of the number of nodes in the communities as long as the number of edges is preserved. Second, modularity has the resolution limit problem that $Q_s$ makes even worse.

The first shortcoming is illustrated in Figure 2 with two simple networks. The left subfigure contains two clique communities and the right subfigure includes two tree communities. In each subfigure, there is one single edge that connects the two communities and there are six edges within all four communities but the number of nodes in clique communities is different from the number of nodes in tree communities. As shown in Table~\ref{tab:simple_example8}, the values of modularity and $Q_s$ of those two different community structures are the same. However, it is quite obvious that the two clique communities have better community structure quality than the two tree communities in terms of node connections. Moreover, this example shows that the number of nodes of the network and within the communities influences neither modularity nor $Q_s$.

Second shortcoming, the resolution limit problem, is illustrated in Figure~\ref{30_cliques}. It displays a ring network comprised of thirty identical cliques, each of which has five nodes and they are connected by single edges. In this case, the modularity of the community structure with each clique forming a different community, totally thirty communities, should be larger than that of the community structure in which two consecutive cliques form a different community, totally fifteen communities. However, Table~\ref{tab:simple_example9} shows that the relation is reversed since the community structure with fifteen communities has larger modularity than that of the community structure with thirty communities. Further, as pointed out in \cite{ResolutionLimit}, when $m(m-1)+2<n$, where $n$ is the number of cliques and $m$ is the number of nodes in each clique, modularity is higher for the large community with two consecutive cliques instead of the small community with a single clique. Moreover, Table~\ref{tab:simple_example9} demonstrates that the difference of $Q_s$ for these two community structures is larger than the corresponding difference of modularity. More specifically, $\Delta Q_s=(0.8424-0.7848)=0.0576>\Delta Q=(0.8879-0.8758)=0.0121$, which means that $Q_s$ makes the resolution limit problem even worse.

To address the above two shortcomings, it is quite intuitive to introduce community density into modularity, incorporating both the number of edges and the number of nodes in the communities and also \textit{Split Penalty}. The corresponding new metric is called \textit{Modularity Density} ($Q_{ds}$). Table~\ref{tab:simple_example8} implies that the $Q_{ds}$ of the two tree communities is almost half of the $Q_{ds}$ of the two clique communities. Moreover, Table~\ref{tab:simple_example9} shows that the $Q_{ds}$ of the community structure in which two consecutive cliques form a different community is almost half of the $Q_{ds}$ of the alternative in which each clique forms a different community. Hence, in this case, $Q_{ds}$ avoids the resolution limit problem. Furthermore, Tables \ref{tab:simple_example1}-\ref{tab:simple_example7} and Figure~\ref{fig_sp_examples} demonstrate that $Q_{ds}$ correctly measures the quality of the community structures of all seven network examples. Even for the network example of Figure~\ref{fig_sp_examples:subfig:d} in which there is ambiguity which community structure is of higher quality, the $Q_{ds}$ of the one big community is only slightly larger than the $Q_{ds}$ of the two small communities as shown in Table~\ref{tab:simple_example4}.

\subsection{Modularity Density}
\label{subsec:modularity_density}
In this subsection, we will give the formulas for $Q_{ds}$ for different kinds of networks, including unweighted and undirected networks, weighted networks, and directed networks, based on the corresponding formulas of $Q_s$ presented in Subsection~\ref{sec:modularity_density}-\ref{subsec:modularity_sp}.

For undirected networks, regardless whether unweighted or weighted, we define $Q_{ds}$ using Equation~(\ref{eq:uwdqs}) as follows
\begin{equation}
\label{eq:uwdqds}
\begin{split}
&Q_{ds}=\sum_{c_i \in C} \biggl[\frac{|E_{c_i}^{in}|}{|E|}d_{c_i}-\left(\frac{2|E_{c_i}^{in}|+|E_{c_i}^{out}|}{2|E|}d_{c_i}\right)^2 \\
& ~~~~~~~~~~~~~-\sum_{\substack{c_j \in C \\ c_j \ne c_i}}\frac{|E_{c_i,c_j}|}{2|E|}d_{c_i,c_j}\biggr], \\
& d_{c_i}=\frac{2|E_{c_i}^{in}|}{|c_i|(|c_i|-1)}, \\
& d_{c_i,c_j}=\frac{|E_{c_i,c_j}|}{|c_i||c_j|}.
\end{split}
\end{equation}
In the above, $d_{c_i}$ is the internal density of community $c_i$, $d_{c_i,c_j}$ is the pair-wise density between community $c_i$ and community $c_j$. Note that $|E_{c_i}^{in}|$ in $d_{c_i}$ and $|E_{c_i,c_j}|$ in $d_{c_i,c_j}$ are unweighted for both unweighted and weighted networks, so that those two community densities are always less than or equal to 1.0.

For directed networks, using Equation~(\ref{eq:directedqs}), $Q_{ds}$ is given by
\begin{equation}
\label{eq:directedqds}
\begin{split}
&Q_{ds}=\sum_{c_i \in C} \biggl[\frac{|E_{c_i}^{in}|}{|E|}d_{c_i} \\
&~~-\frac{(|E_{c_i}^{in}|+|E_{out,c_i}|)(|E_{c_i}^{in}|+|E_{c_i,out}|)}{|E|^2}d_{c_i}^2 \\
&~~-\sum_{\substack{c_j \in C \\ c_j \ne c_i}}\frac{|E_{c_i,c_j}|}{|E|}d_{c_i,c_j}\biggr], \\
& d_{c_i}=\frac{|E_{c_i}^{in}|}{|c_i|(|c_i|-1)}, \\
& d_{c_i,c_j}=\frac{|E_{c_i,c_j}|}{|c_i||c_j|}.
\end{split}
\end{equation}

\section{Evaluation and Analysis}
\label{sec:evaluation}
In this section, we first prove that \textit{Modularity Density} ($Q_{ds}$) solves the resolution limit problem. Then, we introduce two real dynamic datasets and various other popular community quality measurements. Finally, we show the experimental results that validate $Q_{ds}$ ability to solve the two problems of modularity ($Q$) simultaneously.

\subsection{Proof of Solving Resolution Limit Problem}
\label{subsec:resolution_limit_proof}
In this subsection, we test \textit{Modularity Density} ($Q_{ds}$) on the examples from Fortunato and Barth\'{e}lemy \cite{ResolutionLimit}. First, we prove that $Q_{ds}$ does not divide a clique into two or more parts. Then, we verify that $Q_{ds}$ will not merge two or more adjacent cliques connected with a single edge. Finally, we prove that $Q_{ds}$ can discover communities with different sizes.

\textbf{\textit{Modularity Density} ($Q_{ds}$) does not divide a clique into two or more parts.} Given a clique with $m$ ($m \ge 3$) nodes, we prove that maximizing $Q_{ds}$ does not divide this clique into two parts. Consider an arbitrary partition $P$ that divides the clique into communities $c_1$ and $c_2$ with the number of nodes $m_1$ and $m_2$, respectively. Then, the number of edges between $c_1$ and $c_2$ is $m_1m_2$. Let $Q_{ds}(single)$ be the $Q_{ds}$ of the whole clique and $Q_{ds}(pairs)$ be the $Q_{ds}$ of partition $P$. By definitions,
\[
\begin{split}
& Q_{ds}(single)=0, \\
& Q_{ds}(pairs)=\frac{(m_1-m_2)^2-m}{m(m-1)}-\frac{m_1^2+m_2^2}{m^2},
\end{split}
\]
then,
\[
Q_{ds}(pairs)-Q_{ds}(single)=\frac{-2m_1m_2-2m_1m_2m}{m^2(m-1)}<0.
\]
Hence, $Q_{ds}$ will not divide a clique into two parts. A simple generalization of this proof demonstrates that $Q_{ds}$ will not divide a clique into three or more parts.

\begin{figure*}[!t]
\centering
\setlength{\belowcaptionskip}{-1em}
\includegraphics[scale=0.72]{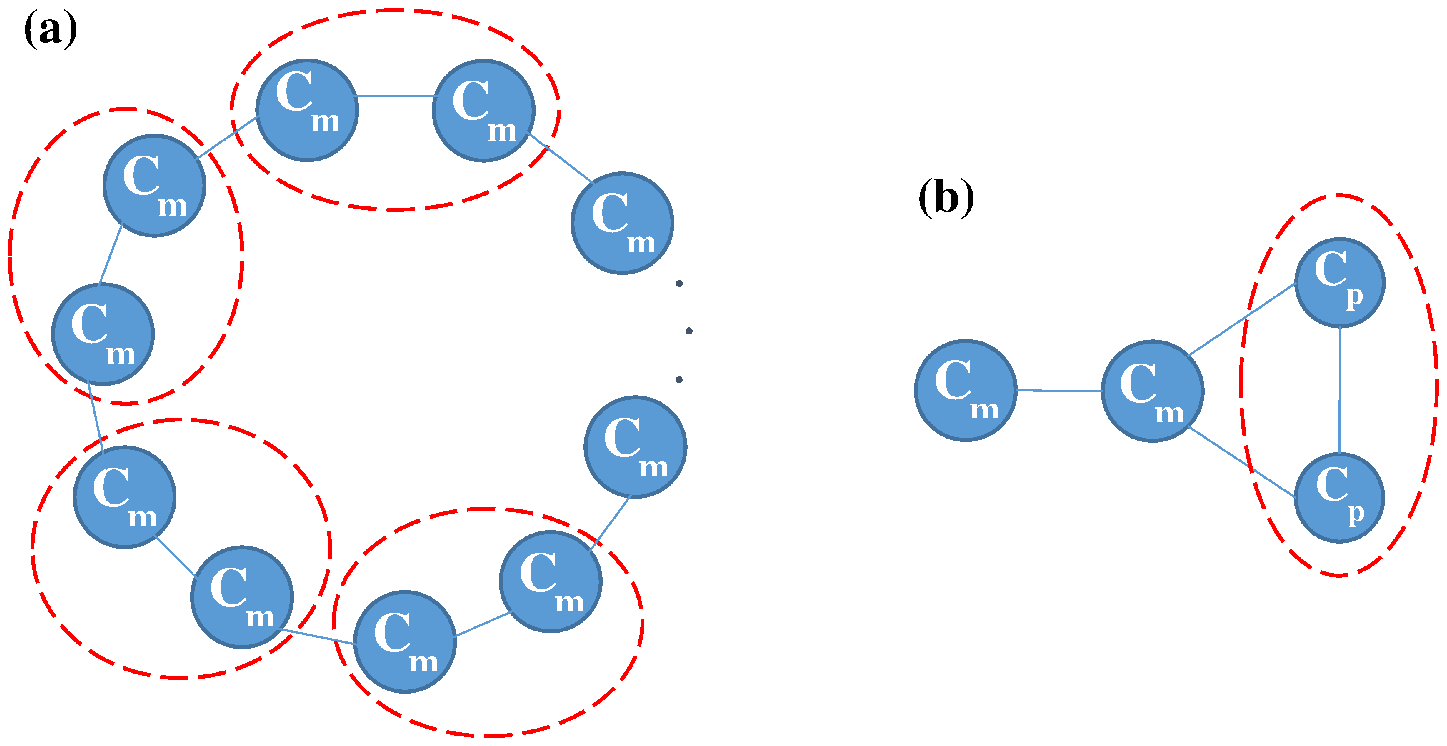}
\vspace{-1.2em}
\centering
\caption{Two clique structure network examples. (a) A clique structure ring network. There are totally $n$ (where $n$ is an even positive integer) cliques. Each clique contains $m$ ($m \ge 3$) nodes, and two consecutive cliques are connected by a single edge. (b) A network with two pairs of identical cliques. One pair of cliques have $m$ ($m \ge 4$) nodes, and the other pair of cliques have $p$ ($3 \le p < m$) nodes.}
\label{clique_resolution}
\vspace{0.9em}
\end{figure*}

\textbf{\textit{Modularity Density} ($Q_{ds}$) does not merge two or more consecutive cliques in the clique structure ring network.} Given a network, see Figure~\ref{clique_resolution}(a), comprised of a ring of $n$ (where $n \ge 2$ is an even integer) cliques connected through single edges. Each clique is a complete graph with $m$ ($m \ge 3$) nodes and $m(m-1)/2$ edges. Then, the cycle network has a total of $nm$ nodes and $nm(m-1)/2+n$ edges. It is clear that the ring network has a well-formed community structure where each community corresponds to a single clique. However, this community structure cannot be obtained by maximizing modularity \cite{ResolutionLimit} since the community structure with $n/2$ communities of two adjacent cliques each has higher modularity. We prove that maximizing $Q_{ds}$ finds the right community structure. We let $Q_{ds}(single)$ be the $Q_{ds}$ of the community structure in which each clique is a different community, totally $n$ communities, and $Q_{ds}(pairs)$ be the $Q_{ds}$ of the community structure with two consecutive cliques forming a different community, totally $n/2$ communities. By definitions,
\[
\begin{split}
&Q_{ds}(single)=\frac{m(m-1)}{m(m-1)+2}-\frac{1}{n}-\frac{2}{m^3(m-1)+2m^2}, \\
&Q_{ds}(pairs)=\frac{\left[m(m-1)+1\right]^2}{\left[m(m-1)+2\right]\left[m(2m-1)\right]} \\
&~~~~~~~~~~~~~~-\frac{2\left[m(m-1)+1\right]^2}{n\left[m(2m-1)\right]^2}-\frac{1}{4m^3(m-1)+8m^2}.
\end{split}
\]
We need to prove the inequality
\begin{equation}
\label{eq:ineq1}
Q_{ds}(pairs)<Q_{ds}(single).
\end{equation}
The first term of $Q_{ds}(pairs)$ can be rewritten as
\[
\frac{[m(m-1)+1]^2}{[m(m-1)+2][m(2m-1)]}=\frac{m^4-2m^3+3m^2-2m+1}{m(m^2-m+2)(2m-1)}.
\]
Then, the first and third terms of $Q_{ds}(single)$ with the latter combined with the last term of $Q_{ds}(pairs)$ yield
\[
-\frac{m^2-m}{m^2-m+2}+\frac{7}{4m^2(m^2-m+2)}=-\frac{m^4-m^3-1.75}{m^2(m^2-m+2)}.
\]
Combining all these terms, we get
\[
\frac{-m^5+m^4+2m^3-2m^2+4.5m-1.75}{m^2(m^2-m+2)(2m-1)}.
\]
We move the remaining two terms to the right hand side of Inequality~(\ref{eq:ineq1}) that we are proving getting
\[
\frac{1}{n}\frac{-2m^4+5m^2-4m+2}{m^2(2m-1)^2}.
\]
Multiplying both sides by $-m^2(2m-1)$ (and changing direction of inequality) we get
\[
\begin{split}
&\frac{m^5-m^4-2m^3+2m^2-4.5m+1.75}{m^2-m+2} \\
&>\frac{1}{4n}\frac{m^4-2.5m^2+2m-1}{m-0.5}.
\end{split}
\]
By doing divisions on both sides, we get
\[
\begin{split}
&m^3-4m-2+\frac{1.5m+5.75}{m^2-m+2} \\
&>\frac{1}{4n}\left[m^3+0.5m^2-2.25m+0.875-\frac{9}{16m-8}\right].
\end{split}
\]

Since $\frac{1.5m+5.75}{m^2-m+2}\geq 0$, and $\frac{1}{4n}\leq\frac{1}{8}$ for $n\geq 2$ and also $\frac{9}{16m-8}>0$, we just need to show that
\[
m^3-4m-2>\frac{m^3+0.5m^2-2.25m+0.875}{8}
\]
which simplifies to
\[
7m^3-0.5m^2-29.75m-16.875>0 \mbox{ for } m\geq 3,
\]
which is easy to prove either by induction, starting at $m=3$, or by inspecting zeros of the derivative $21m^2-m-29.75$, which are all less than $2.0$, showing that this polynomial is positive for $m\geq 3$.

Since Inequality (\ref{eq:ineq1}) holds, $Q_{ds}$ will not merge two consecutive cliques in the ring network. A straightforward extension of the proof shows that $Q_{ds}$ will not merge three or more consecutive cliques.

\textbf{\textit{Modularity Density} ($Q_{ds}$) could discover communities with different sizes.} Consider a network, shown in Figure~\ref{clique_resolution}(b), with two pairs of identical cliques. The left pair of cliques have $m$ ($m \ge 4$) nodes, and the right pair of cliques have $p$ ($3 \le p < m$) nodes. This network has $2m+2p$ nodes and $m(m-1)+p(p-1)+4$ edges. It is obvious that each of the four cliques should be a different community. However, the authors in \cite{ResolutionLimit} found that maximizing modularity will merge the right two small cliques. Here, we prove that maximizing $Q_{ds}$ will not merge them. We let $Q_{ds}(single)$ denote the $Q_{ds}$ of the community structure in which each clique corresponds to a single clique, and $Q_{ds}(pairs)$ be the $Q_{ds}$ of the community structure with the right two small cliques merged into one community. Clearly, the $Q_{ds}$ of the left two large cliques will stay the same in those two different community structures so we denote it as $Q_{ds}(0)$. By definitions,
\[
\begin{split}
&Q_{ds}(single)=Q_{ds}(0)+\frac{p(p-1)}{m(m-1)+p(p-1)+4} \\
&~~~~~~~~~-\frac{\left[p(p-1)+2\right]^2}{2\left[m(m-1)+p(p-1)+4\right]^2} \\
&~~~~~~~~~-\frac{1}{mp\left[m(m-1)+p(p-1)+4\right]} \\
&~~~~~~~~~-\frac{1}{p^2\left[m(m-1)+p(p-1)+4\right]}, \\
\end{split}
\]
\[
\begin{split}
&Q_{ds}(pairs)=Q_{ds}(0)-\frac{1}{mp\left[m(m-1)+p(p-1)+4\right]} \\
&~~~~~~~~~-\frac{\left[p(p-1)+1\right]^2\left[p(p-1)+2\right]^2}{p^2(2p-1)^2\left[m(m-1)+p(p-1)+4\right]^2} \\
&~~~~~~~~~+\frac{\left[p(p-1)+1\right]^2}{p(2p-1)\left[m(m-1)+p(p-1)+4\right]}.
\end{split}
\]

The inequality that we need to prove is
\begin{equation}
\label{eq:ineq2}
Q_{ds}(single)-Q_{ds}(pairs)>0.
\end{equation}
Since
\[
\begin{split}
&Q_{ds}(single)-Q_{ds}(pairs) \\
&~=\frac{1}{m(m-1)+p(p-1)+4}*\biggl\{ p(p-1)-\frac{1}{p^2}\\
&~-\frac{[p(p-1)+2]^2}{2[m(m-1)+p(p-1)+4]}-\frac{[p(p-1)+1]^2}{p(2p-1)} \\
&~+\frac{[p(p-1)+1]^2[p(p-1)+2]^2}{p^2(2p-1)^2[m(m-1)+p(p-1)+4]}
\biggr\},
\end{split}
\]
it is clear that the first factor is always positive so it can be removed from consideration and the interior of the second factor can be rewritten as
\[
(p^2-p)+
\]
\[
\frac{2[p^2-p+1]^2[p^2-p+2]^2-[p^2-p+2]^2p^2(2p-1)^2}{2p^2(2p-1)^2[m^2-m+p^2-p+4]}
\]
\[
>\frac{1}{p^2}+\frac{[p^2-p+1]^2}{p(2p-1)}.
\]
The second term simplifies to
\[
-\frac{[p^2-p+2]^2}{2}\frac{2p^4-5p^2+4p-2}{p^2(2p-1)^2[m^2-m+p^2-p+4]}.
\]
Since by induction for $p\geq 3$ the polynomial $2p^4-5p^2+4p-2$ is positive, then this term is greater than
\[
\begin{split}
&-\frac{(p^2-p+2)(2p^4-5p^2+4p-2)}{4p^2(2p-1)^2} \\
&=\frac{1}{4}\left[-0.5p^2+0.375-\frac{7.5p^3-15.625p^2+10p-4}{4p^4-4p^3+p^2}\right].
\end{split}
\]
It is easy to show that the last fraction is less than 0.391 by using induction or by finding zeros of the fraction derivative, which are all less than 2.5, so we just need to prove that $0.875p^2-p-0.004$ is greater than the right hand side of Inequality (\ref{eq:ineq2}).

The second term of the right hand side of Inequality (\ref{eq:ineq2}) can be rewritten as
\[
\begin{split}
&\frac{p^4-2p^3+3p^2-2p+1}{2(p^2-0.5p)}=0.5p^2-0.75p+1.125\\
&-\frac{0.875p-1}{2(p^2-0.5p)}<0.5p^2-0.75p+1.125,
\end{split}
\]
because $0.875p>1$ and $p^2>p$ for $p\geq 2$.

Since $\frac{1}{p^2}<0.12$, the inequality that we need to prove reduces to
$0.375p^2-0.25p>1.249$, but for $p\geq 3$, $0.375p^2-0.25p\geq 2.625$, proving Inequality (\ref{eq:ineq2}). Thus, we conclude that maximizing $Q_{ds}$ will not merge the right two small cliques, demonstrating that $Q_{ds}$ can discover communities of different size.

In summary, all the above proofs show that \textit{Modularity Density} solves the resolution limit problem of modularity.

\subsection{Real Dynamic Datasets}
\label{subsec:datasets}
In this subsection, we introduce two real dynamic datasets on which we conduct experiments in order to validate $Q_{ds}$ avoids the two problems of modularity.

\textbf{Senate Dataset} \cite{Estrangement,SenateDataset}. The Senate dataset is a time-evolving weighted network comprised of United States senators where the weight of an edge represents the similarity of their roll call voting behavior. This dataset was obtained from website \textit{voteview.com} and the similarities between a pair of senators were calculated following Waugh et al. \cite{SenateDataset} as the number of bills for which the senators of the pair voted the same way, normalized by the number of bills for which they both voted. The dataset totally consists of $111$ snapshots corresponding to Senate's activities over $220$ years and includes $1916$ unique senators.

\textbf{Reality Mining Bluetooth Scan Data} \cite{RealityMining}. This dataset was created from the records of Bluetooth Scans generated among the $94$ subjects in Reality Mining study conducted from 2004-2005 at the MIT Media Laboratory. In the network, nodes represent the subjects and the directed edges correspond to the Bluetooth Scan records while the weight of each edge represents the number of direct Bluetooth scans between the two subjects. In the experiments, we only used the records from August 02, 2004 (Monday) to May 29, 2005 (Sunday) and we divided them into weekly snapshots, so each snapshot represents scans collected during the corresponding week. There are total of 43 snapshots.

\begin{table*}[!t]
\vspace{1em}
\caption{The average metric differences between LabelRankT with different values of conditional update parameter $q$ and Estrangement on Senate dataset.}
\label{senate}
\vspace{0.1em}
\centering
\setlength{\tabcolsep}{0.4pt}
\begin{tabular}{c|p{0.5pt}|c|c|c|c|c|c|c|c|c|c|c}
\hline \hline
     LabelRankT $q$ && 0.05 & 0.1 & 0.2 & 0.3 & 0.4 & 0.5 & 0.6 & 0.7 & 0.8 & 0.9 & 0.95 \\
\hline
     $Q$ && -0.0534 & -0.0462 & \textbf{-0.0408} & -0.0538 & -0.0714 & -0.0848 & -0.083 & -0.0897 & -0.0897 & -0.0848 & -0.08 \\
\hline
     $Q_s$ && -0.166 & -0.0802 &	0.0468 & 0.0808 & 0.0969 & 0.112 & \textbf{0.116} & 0.115 & 0.115 & 0.111 & 0.106 \\
\hline
     $Q_{ds}$ && -0.1638 & -0.0787 & 0.04847 & 0.08297 & 0.0995 & 0.1145 & 0.1182 &	\textbf{0.1183} & \textbf{0.1183} & 0.1135 & 0.1083 \\
\hline
    \# \textit{Intra-edges} && -159.102 & -32.444 & 234.296 & 387.38 & 510.645 & 616.855 & 615.123 & \textbf{624.764} & \textbf{624.764} & 602.627 & 580.733 \\
\hline
     \textit{Contraction} && -6.806 & -3.023 & 2.481 & 4.553 & 5.937 & 7.033 & 7.065 & \textbf{7.227} & \textbf{7.227} & 6.927 & 6.622 \\
\hline
    \# \textit{Inter-edges} && -75.962 & -54.098 & -123.898 & -187.99 & -245.198 & -299.356 &	-300.108 & \textbf{-303.043} & \textbf{-303.043} & -292.782 &	-282.442 \\
\hline
     \textit{Expansion} && 6.448 & 2.91 &	-2.428 & -4.416 & -5.737 & -6.847 & -6.878 & \textbf{-7.009} & \textbf{-7.009} & -6.724 & -6.431 \\
\hline
     \textit{Conductance} && 0.213 & 0.0851 & -0.0886 & -0.148 & -0.186 &	-0.214 & -0.216 & \textbf{-0.224} & \textbf{-0.224} & -0.213 & -0.201 \\
\hline \hline
\end{tabular}
\vspace{-0.5em}
\end{table*}

\begin{table*}[!t]
\caption{The average metric differences between LabelRankT with different values of conditional update parameter $q$ and Estrangement on reality mining bluetooth scan data.}
\label{reality_mining}
\vspace{0.1em}
\centering
\setlength{\tabcolsep}{0.1pt}
\begin{tabular}{c|p{1.2pt}|c|c|c|c|c|c|c|c|c|c|c}
\hline \hline
     LabelRankT $q$ && 0.05 & 0.1 & 0.2 & 0.3 & 0.4 & 0.5 & 0.6 & 0.7 & 0.8 & 0.9 & 0.95 \\
\hline
     $Q$ && -0.161 & -0.121 & -0.0783 & -0.0744 & -0.0724 & \textbf{-0.0699} & -0.0702 & -0.0724 & -0.0742 & -0.0755 & -0.0774 \\
\hline
     $Q_s$ && -0.379 & -0.244 & -0.107 & -0.0802 & -0.0538 & -0.0497 & \textbf{-0.0382} & -0.0405 & -0.0521 & -0.0634 & -0.0713 \\
\hline
     $Q_{ds}$ && -0.191 & -0.0984 & -0.0222 & -0.017 & -0.0116 & -0.0116 & \textbf{-0.00318} & -0.00826 & -0.011 & -0.0115 & -0.0134 \\
\hline
    \# \textit{Intra-edges} && -1450.893 & -956.006 & -479.377 & -331.371 & -230.263 & -183.536 & -102.94 & \textbf{-78.93} & -155.183 & -242.287 & -333.419 \\
\hline
     \textit{Contraction} && -86.909 & -69.914 & -52.543 & -46.371 & -43.176 & -40.567 & \textbf{-35.948} & -36.425 & -38.006 & -41.277 & -45.425 \\
\hline
    \# \textit{Inter-edges} && -39.949 & -76.524 & -159.74 & -167.333 & -190.947 & -190.865 & \textbf{-196.098} & -193.123 & -188.708 & -179.653 & -178.96 \\
\hline
     \textit{Expansion} && 52.529 & 25.829 & 6.289 & 5.76 & 5.664 & 7.07 & \textbf{4.881} & 6.799 & 6.916 & 6.117 & 5.669 \\
\hline
     \textit{Conductance} && 0.23 & 0.176 & 0.114 & 0.1 & 0.0934 & 0.0933 & \textbf{0.0843} & 0.0955 &	0.102 &	0.107 &	0.104 \\
\hline \hline
\end{tabular}
\vspace{0.3em}
\end{table*}

\subsection{Community Quality Measurements}
\label{subsec:metrics}
In the discussion of the experimental results we use various community quality metrics, including the number of \textit{Intra-edges}, \textit{Contraction}, the number of \textit{Inter-edges}, \textit{Expansion}, and \textit{Conductance} \cite{CommunityMetrics}, which characterize how community-like is the connectivity structure of a given set of nodes. All of them rely on the intuition that communities are sets of nodes with many edges inside them and few edges outside of them. Now, given a network $G=(V,E)$ and given a community or a set of nodes $c$, let $|c|$ be the number of nodes in the community $c$ and let $|E_c^{in}|$ denote the total number of edges in $c$ for unweighted networks or the total weight of such edges for weighted networks. We denote the total number of edges from the nodes in community $c$ to the nodes outside $c$ for unweighted networks or the total weight of such edges for weighted networks as $|E_c^{out}|$. Then, the definitions of the five quality metrics are as follows: \\
\textbf{The number of \textit{Intra-edges}:} $|E_c^{in}|$; it is the total number of edges in $c$ or the total weight of such edges. A large value of this metric is better than a small value in terms of the community quality. \\
\textbf{\textit{Contraction}:} $2|E_c^{in}|/|c|$ for undirected networks or $|E_c^{in}|/|c|$ for directed networks; it measures the average number of edges per node inside the community $c$ or the average weight per node of such edges. A large value of \textit{Contraction} is better than a small value in terms of the community quality. \\
\textbf{The number of \textit{Inter-edges}:} $|E_c^{out}|$; it is the total number of edges from the nodes in community $c$ to the nodes outside $c$ or the total weight of such edges. A small value of this metric is better than a large value in terms of the community quality.  \\
\textbf{\textit{Expansion}:} $|E_c^{out}|/|c|$; it measures the average number of edges (per node) that point outside the community $c$ or the average weight per node of such edges. A small value of \textit{Expansion} is better than a large value in terms of the community quality. \\
\textbf{\textit{Conductance}:} $\frac{|E_c^{out}|}{2|E_c^{in}|+|E_c^{out}|}$ for undirected networks or $\frac{|E_c^{out}|}{|E_c^{in}|+|E_c^{out}|}$ for directed networks; it measures the fraction of the total number of edges that point outside the community for unweighted networks or the fraction of the total weight of such edges for weighted networks. A small value of \textit{Conductance} is better than a large value in terms of the community quality.

\begin{figure*}[!t]
\centering
\setlength{\belowcaptionskip}{-1em}
\subfigure[Senate dataset ($q=0.7$).]{
\label{exp_q:subfig:a}
\includegraphics[scale=0.49]{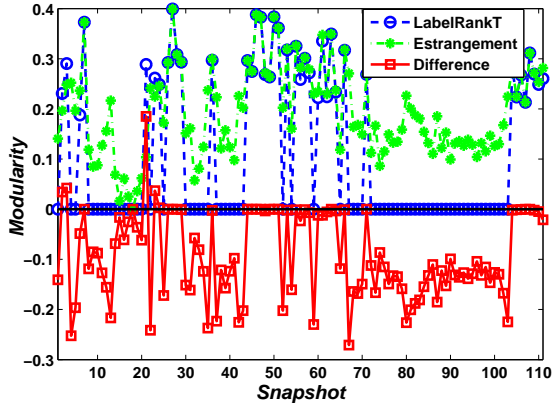}
}
\hspace{1.5em}
\subfigure[Reality Mining Bluetooth Scan data ($q=0.6$).]{
\label{exp_q:subfig:b}
\includegraphics[scale=0.49]{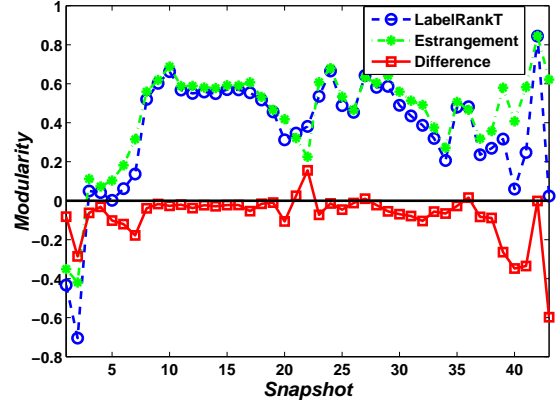}
}
\vspace{-1.4em}
\centering
\caption{The modularity ($Q$) of the community detection results of LabelRankT and Estrangement (also, the difference between LabelRankT and Estrangement) on (a) each snapshot of Senate dataset at $q=0.7$ and on (b) each snapshot of Reality Mining Bluetooth Scan data with $q=0.6$.}
\label{exp_q}
\vspace{0.8em}
\end{figure*}

\subsection{Experimental Results}
\label{subsec:experiment}
In this subsection, we report the results of performing community detection on the two real dynamic datasets introduced in Subsection~\ref{sec:evaluation}-\ref{subsec:datasets} by using the dynamic community detection algorithms, LabelRankT \cite{LabelRankT} and Estrangement \cite{Estrangement}. We chose these two algorithms because the second algorithm relies on the modularity optimization while the first one does not. In the experiments, we adopted the best parameter of Estrangement but varying the conditional update parameter $q \in [0,1]$ of LabelRankT from 0.05 to 0.95. As seen in the results, in most cases, the best $q$ is around 0.7 in agreement with the best value reported in \cite{LabelRankT}. For the community structures found by the two algorithms, we calculated the values of modularity ($Q$), $Q_s$, \textit{Modularity Density} ($Q_{ds}$), and the five community quality metrics described in Subsection~\ref{sec:evaluation}-\ref{subsec:metrics}.

\begin{figure*}[!t]
\centering
\setlength{\belowcaptionskip}{-1em}
\subfigure[Senate dataset ($q=0.7$).]{
\label{exp_qs:subfig:a}
\includegraphics[scale=0.49]{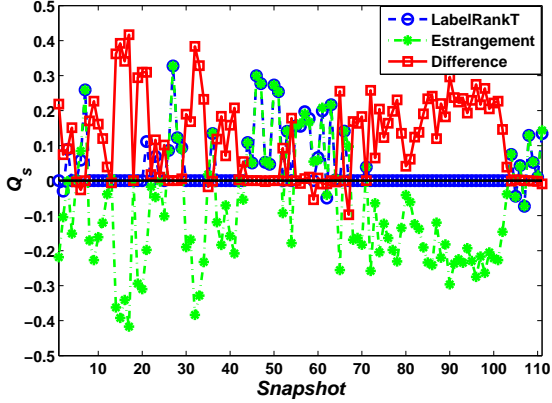}
}
\hspace{1.5em}
\subfigure[Reality Mining Bluetooth Scan data ($q=0.6$).]{
\label{exp_qs:subfig:b}
\includegraphics[scale=0.49]{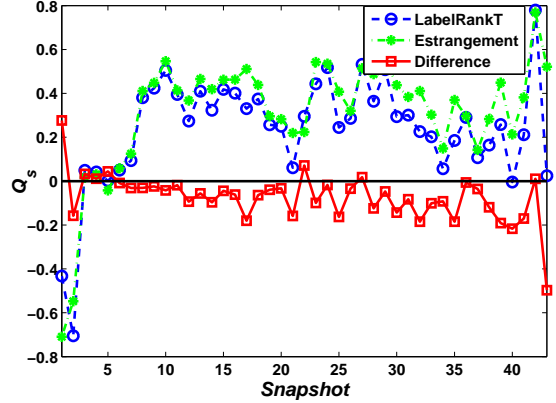}
}
\vspace{-1.4em}
\centering
\caption{$Q_s$ of the community detection results of LabelRankT and Estrangement (also, the difference between LabelRankT and Estrangement) on (a) each snapshot of Senate dataset at $q=0.7$ and on (b) each snapshot of Reality Mining Bluetooth Scan data with $q=0.6$.}
\label{exp_qs}
\vspace{-1em}
\end{figure*}

\begin{figure*}[!t]
\vspace{2em}
\centering
\setlength{\belowcaptionskip}{-1em}
\subfigure[Senate dataset ($q=0.7$).]{
\label{exp_qds:subfig:a}
\includegraphics[scale=0.49]{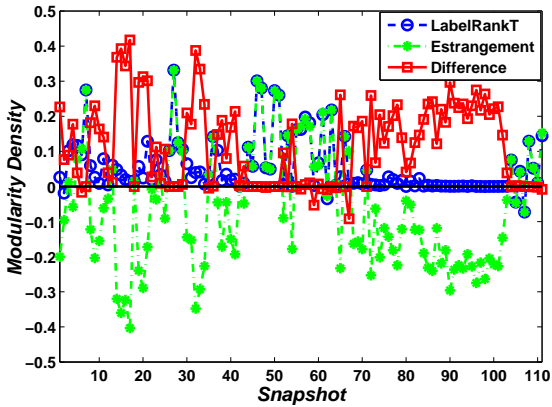}
}
\hspace{1.5em}
\subfigure[Reality Mining Bluetooth Scan data ($q=0.6$).]{
\label{exp_qds:subfig:b}
\includegraphics[scale=0.49]{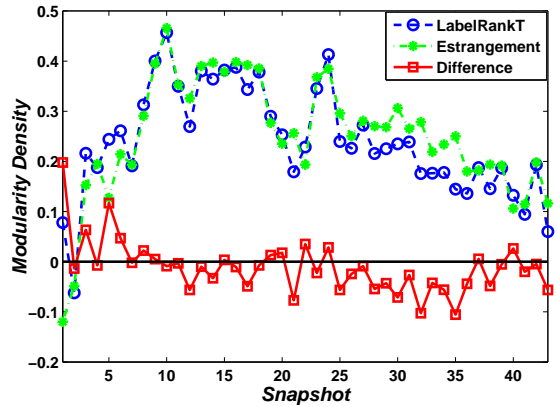}
}
\vspace{-1.4em}
\centering
\caption{The \textit{Modularity Density} ($Q_{ds}$) of the community detection results of LabelRankT and Estrangement (also, the difference between LabelRankT and Estrangement) on (a) each snapshot of Senate dataset at $q=0.7$ and on (b) each snapshot of Reality Mining Bluetooth Scan data with $q=0.6$.}
\label{exp_qds}
\vspace{0.8em}
\end{figure*}

Table~\ref{senate} and Table~\ref{reality_mining} present the average metric differences between LabelRankT with different values of conditional update parameter $q$ and Estrangement on Senate dataset and Reality Mining Bluetooth Scan data, respectively. That is, we first computed the values of the eight metrics above for the community detection results, detected by Estrangement, of each snapshot. Then, we calculated the eight metrics values for the community detection results, discovered by LabelRankT for all $q$, of each snapshot. Next, we got the metric differences of all eight metrics by subtracting the metric values of Estrangement from those of LabelRankT for all $q$'s over each snapshot. Then, averaging those differences of each metric over all the snapshots, we obtained the corresponding average metric differences.

Table~\ref{senate} demonstrates that $Q$ gets its largest value when $q=0.2$; $Q_s$ reaches the largest value when $q=0.6$; $Q_{ds}$, \textit{Intra-edges}, and \textit{Contraction} get their largest values at $q=0.7$ and $q=0.8$; also, \textit{Inter-edges}, \textit{Expansion}, and \textit{Conductance} reach their smallest values at $q=0.7$ and $q=0.8$. Thus, $Q_{ds}$ is consistent with the five metrics introduced in Subsection~\ref{sec:evaluation}-\ref{subsec:metrics} on determining the best $q$ for LabelRankT on Senate dataset while $Q$ and $Q_s$ are not consistent with them. Further, we could observe that $Q$ is always negative which indicates that LabelRankT performs below Estrangement over all $q$'s because the goal of Estrangement is to maximize modularity ($Q$). However, the other seven metrics imply that LabelRankT performs better than Estrangement when $q>0.1$. Therefore, we could explicitly observe that maximizing $Q$ to detect communities has problems in measuring the community detection quality correctly on Senate dataset.

Table~\ref{reality_mining} shows that six metrics get their best (largest or smallest) values at $q=0.6$ while the two exceptions, $Q$ and the number of  \textit{Intra-edges}, reach their largest values when $q=0.5$ and $q=0.7$, respectively. Thus, the six metrics, except $Q$ and the number of \textit{Intra-edges}, are consistent on determining the best value of $q$ for LabelRankT on Reality Mining Bluetooth Scan data. This indicates that on Reality Mining Bluetooth Scan data, maximizing $Q$ to detect communities has problems.

It is also interesting to observe that for $q=0.05$ and $q=0.1$ in Table~\ref{senate}, \textit{Inter-edges} metric implies that LabelRankT performs better than Estrangement on Senate dataset, which is not consistent with $Q_s$, $Q_{ds}$, \textit{Intra-edges}, \textit{Contraction}, \textit{Expansion}, and \textit{Conductance} metrics. Moreover, we could learn from Table~\ref{reality_mining} that all the metrics, except \textit{Inter-edges} metric, imply that LabelRankT performs slightly below the performance of Estrangement over all $q$'s.  Thus, \textit{Inter-edges} metric has some problems. Also, as mentioned in the paragraph above, \textit{Intra-edges} metric is not consistent with the other six metrics on determining the best $q$ for LabelRankT, which also means that \textit{Intra-edges} metric has problems. We conjecture that the reason for the shortcoming of \textit{Intra-edges} and \textit{Inter-edges} metrics is the same as the case of modularity ($Q$) which does not consider the number of nodes in the communities. This reason also implies the superiority of $Q_{ds}$ over $Q$ and $Q_s$.

Based on the results presented in the above two tables, we conclude that $Q_{ds}$ solves the two problems of modularity. We also conjecture that the difference between the best values of $q$ for LabelRankT determined by $Q$ and $Q_{s}$ and the difference determined by $Q_s$ and $Q_{ds}$ on Senate dataset is a manifestation of the two problems of modularity maximization, namely favoring small communities and the resolution limit problem. Moreover, the difference between the best values of $q$ for LabelRankT determined by $Q$ and $Q_{s}$  on Reality Mining Bluetooth Scan data indicates that maximizing $Q$ has the problem of favoring small communities. Thus, $Q_s$ and $Q_{ds}$ can be used for checking whether finding communities by maximizing $Q$ on a specific dataset will suffer any of the two problems.

To make the differences among $Q$, $Q_s$, and $Q_{ds}$ more clear, we plot their values, in Figures~\ref{exp_q},~\ref{exp_qs}, and~\ref{exp_qds}, of the community detection results of LabelRankT and Estrangement on each snapshot of Senate dataset at $q=0.7$ and on each snapshot of Reality Mining Bluetooth Scan data when $q=0.6$. Figure~\ref{exp_q:subfig:a} shows that in most cases $Q$ is negative, while $Q_s$ and $Q_{ds}$ are positive as seen in Figure~\ref{exp_qs:subfig:a} and Figure~\ref{exp_qds:subfig:a}. It indicates that there is large difference between $Q$ and $Q_s$ or between $Q$ and $Q_{ds}$. This is consistent with Table~\ref{senate}. Further, it can be observed from Figure~\ref{exp_qs:subfig:a} and Figure~\ref{exp_qds:subfig:a} that $Q_s$ and $Q_{ds}$ are almost the same on each snapshot, which is also consistent with Table~\ref{senate}. Figure~\ref{exp_q:subfig:b}, Figure~\ref{exp_qs:subfig:b}, and Figure~\ref{exp_qds:subfig:b} demonstrate that $Q$, $Q_s$, and $Q_{ds}$ are negative in most of the cases, although their values are different in each snapshot. These observations are consistent with the results shown in Table~\ref{reality_mining}.

\section{Conclusion and Future Work}
\label{sec:conclusion}
In this paper, we propose a new community quality metric, called \textit{Modularity Density}, which solves the problems of modularity of favoring small communities in some circumstances and large communities in others. We demonstrate with proofs and experiments on real dynamic datasets that \textit{Modularity Density} is an effective alternative to modularity.

In the future, we plan to extend \textit{Modularity Density} to enable evaluation of the quality of overlapping community structures. We will also propose a community detection algorithm based on \textit{Modularity Density} maximization and then compare its community detection results with those of modularity maximization algorithms on some typical real networks.

\section*{Acknowledgment}
This work was supported in part by the Army Research Laboratory under Cooperative Agreement Number W911NF-09-2-0053 and by the the Office of Naval Research Grant No. N00014-09-1-0607. The views and conclusions contained in this document are those of the authors and should not be interpreted as representing the official policies either expressed or implied of the Army Research Laboratory or the U.S. Government.



\end{multicols*}
\end{document}